\documentclass[%
rsi,
 amsmath,amssymb,
 reprint,%
author-numerical,%
]{revtex4-1}

\usepackage{graphicx}
\usepackage{dcolumn}

\usepackage{setspace}
\usepackage{todonotes}
\usepackage{subcaption}
\usepackage{bm}
\usepackage{psfrag}
\usepackage[utf8]{inputenc}
\usepackage[T1]{fontenc}
\usepackage{mathptmx}

\begin{document}

\preprint{AIP/123-QED}

\title[Simultaneous Eulerian-Lagrangian velocity measurements of particulate pipe flow in transitional regime]{Simultaneous Eulerian-Lagrangian velocity measurements of particulate pipe flow in transitional regime}

\author{S. Singh}
\email{singhs71@uni.coventry.ac.uk}

\author{A. Poth\'erat}%

\author{C. C. T. Pringle}

\author{I. R. J. Bates}

\author{ Martin Holdsworth}

\affiliation{ 
Fluid and Complex Systems Centre, Coventry University, CV1 5FB, UK.
}%

\date{\today}

\begin{abstract}
We present a unique pipe flow rig capable where simultaneous particle tracking and flow velocity measurements in a dilute, neutrally buoyant particulate pipe flow in regimes of transition to turbulence. The flow consists of solid glass spheres for the disperse phase, and a density-matching fluid for the carrier phase. The measurements are conducted using a bespoke, combined 2D PIV/ PTV technique. The technique takes advantage of a phase discrimination approach that involves separating the disperse and carrier phases based on their respective image characteristics. Our results show that the rig and the technique it implements can be employed to study transitional particulate pipe flows at dilute concentration.
\end{abstract}

\maketitle

\section{\label{sec:intro}Introduction}
Adding particles, even in moderate amounts to an otherwise Newtonian flow can drastically alter its global properties. For instance, they are commonly added into pipes to reduce drag in various industrial applications \cite{gad2006flow,crowe}. While in some cases, particulate flows seem to transition to turbulence at higher Reynolds numbers than their single phase counterpart, the underlying mechanisms are not understood. One of the reasons is the lack of experiments targeting the transitional regimes, where both fluid and particles can be tracked simultaneously and separately. The purpose of this work is precisely to address this gap, in the canonical case of 
particulate pipe flows and neutrally buoyant particles.\\
Transition in particulate pipe flow is much more complex than in single phase flow in that it is not only determined by the flow rate but also the particles size and concentration. Despite wide applications in industries such as oil and gas or chemical and food processing, it has not 
been studied extensively. 
Early experiments on the transition to turbulence in a pipe highlighted a critical volume fraction of
particles below which they favoured the transition at a lower Reynolds number. At higher volume fractions, by contrast, this effect was reversed \citep{Matas}. Recent numerical simulations based on accurate modelling of individual solid particles recovered this phenomenology for pipe \citep{Yu} and channel flows \citep{wang_richter_2019}. Between very small particles that passively trace the flow and large ones which ignore it, there is an intermediate optimal size of particles which have the greatest effect on the flow. Here, adding only  minimal additional physics is sufficient to greatly alter the linear growth of modal and non-modal perturbations \citep{klinkenberg2011,Klinkenberg2013PRe,anthony_royal, boronin2012optimal} to the point of making pipe flows linearly unstable \cite{anthony}. \\
 Several experiments 
\citep{darbyshire_mullin_1995,Peixinho,nishi_durst_biswas_2008} have provided a 
more quantitative description of the dynamics of transition in single phase pipe flow, and its flow 
structures during the transitional phase \citep{HofDoorne,wygnanski_champagne_1973, wygnanski_sokolov_friedman_1975}. A more detailed review of some of the experiments in pipe flows 
can be found here \citep{annurev-fluid-mullin}. The main feature of the pipe flow problem is the sub-critical nature of its transition to turbulence that makes it highly sensitive to external disturbances. As far as experiments are concerned, extreme care is needed to keep out unwanted perturbations and introduce only
controlled ones, so as to precisely identify the conditions of the transition.
In regards to experiments in particulate pipe flow, \cite{Matas} quantified transition by pressure drop measurements, \cite{Particle_Ultrasound_poelma, Hof_particle} looked at the effects of volume fractions on transition scenarios and identified regimes where transition scenarios were different from that of Newtonian-type. While \citep{Particle_Ultrasound_poelma} measured just the fluid velocity using Ultra sound image velocimetry (UIV), simultaneous measurement of fluid-particle remains elusive for neutrally buoyant, particulate pipe flow in the transitional regimes. These experiments have increased our understanding of transition in particle-laden flow, however, the role played by particles in instigating transition is still not clear. The measurement systems used in these experiments are ``blind'' to particles and effective only at diagnosing the particles' effect on the carrier phase. In regards to the universality of the transition regimes \citep{Particle_Ultrasound_poelma, Hof_particle}, further experimental data is needed with varying particle sizes and different perturbation systems to consider this question closed. Additionally, the regimes considered by \cite{anthony} that are suggestive of possible instability have also not been explored.\\
Measuring the velocities of both the fluid and the particles by optical means require two distinct 
techniques. Using Particle Image Velocimetry (PIV), the instantaneous whole field velocity (Eulerian) of single 
phase can be measured. In particle-laden flow, by contrast, a low image density technique 
such as Particle Tracking Velocimetry (PTV) typically provides the Lagrangian velocity of the disperse 
phase velocity. However, these techniques become quite challenging when implemented simultaneously, and successful implementation would require a trade-off. A moderate concentration of discrete particles leads to an overload of the optical images that incur
restrictions, imposed by concentration of discrete particles \citep{adrian_westerweel_2011}. The difficulty in distinguishing 
the signals from each phase often causes false positive cross correlation between them, also known as ``cross-talk'' \citep{Poelma,brucker2000piv}.\\
To circumvent this issue, a phase discrimination method based on image intensity is implemented to separate the phases by dynamic masking. In this method, image thresholding and image feature selection are used to identify the disperse phase which subsequently generates the disperse phase velocities. For the fluid phase velocity, the identified disperse phase undergoes dynamic masking before the PIV processing techniques. The details of this method are elaborated further in Sect.  \ref{sec:5}.
The main challenge in this approach is to accurately discriminate between the dispersed phase and the 
carrier phase. In the past, this has been tackled in several ways: \cite{Poelma} used fluorescent particle tagging to optically separate the 
signal between the fluorescent tracer and particles; \cite{kiger_pan_2000,Jakobsen_twophase,kieger_sus_exp} and 
\cite{Knowles2012} separated the phases by image processing based on the relative sizes and optical 
scattering of the tracers and particles; \cite{jetparticleladen} expanded on previous work of \cite{Khalitov2002} by implementing versatile image filters and particle size-brightness maps; \cite{Elhimer_flor} used combination of both the methods to 
study turbulent particle-laden flow. In other notable research works, phase separation was achieved by ensemble correlation algorithm \citep{DeenTwoPhase}, by edge detection of bright features \citep{brucker2000piv}, whereas \cite{Kolaas} used erosion and dilation techniques to separate the phases before processing the PIV and PTV using the Digiflow software \citep{Dalziel1992}. Furthermore, generating individual logical masks (dynamic masking) over each frames 
pose difficulties when the particles and tracers have small contrast, and if the uneven illumination causing inhomogenous background cannot be removed with background subtraction \citep{brucker2000piv}. Such issues can be mitigated by evenly illuminating the plane with a thin light sheet and using a shallow depth of field. Logical masks obtained by padding zeros to the particle features can create ``coronos'' from the edges of particles that were not masked properly, resulting in erroneous cross-correlation. This issue can be mitigated by pre-treating the masked images with a median filter \cite{brucker2000piv}. Failing that, one can resort to recent advanced techniques \citep{randommasking,automatedmask}. However, if the particle shape is known \textit{a priori}, masks of same sizes as that of particle detected can be implemented without implementing computationally costly algorithm. One of the inherent limitations of PIV techniques is that at high particles volume fraction ($>10^{-3}$), obscuration prevents optical measurements. This can be overcome by carefully matching the refractive indices of solid and the carrier phases \citep{Wiederseiner2011}, and implementing in refractive index matching (RIM) PIV \citep{zade}. However, there are limitations posed by the PTV tracking algorithm which cannot identify particles correctly anymore \citep{adrian_westerweel_2011}. Additionally, with more solid particles detected and masked, higher regions of missing vectors in the fluid phase cause significant data loss \citep{DeenTwoPhase}, interpolation of which results in spurious nonphysical results \citep{Poelma}.\\
While there are several particle tracking methods \citep{nature_tracking,Brevis2011,SBALZARINI}, each of them have their own strengths and limitations in regards to their particle tracking capabilities, implementation and computational costs. Common strategy in tracking algorithm is to look for the nearest neighbour in a search window \citep{CROCKER}, however the data yield in such algorithm is poor since it ends particle tracks in the event of an overlap between particles. Multi-frame tracking \citep{Ouellette2006,Kelley} performs well in such situations by starting new tracks. However, tracking still remains challenging in 2D PIV where tracks are inherently short because particles move in and out of a thin LASER sheet. A 3D PTV system such as the one used in turbulent pipe flow measurements by \cite{Olivera3D}, or one with advanced tracking features based on neural network techniques \citep{Shakethebox} would be able to reconstruct short, broken tracks; however, it would still suffer from data loss due to masking. Futhermore, there are no measurement systems yet that can simultaneously perform full Lagrangian tracking of particles and solids, and is available in numerical modelling \cite{maxey_review}.\\
The vast majority of research work in this area shows clearly how complex these techniques are, and that there are no generic solutions that would fit and reconcile all experimental difficulties. Alternatives to PIV techniques to diagnose high volume fraction ``opaque" flows exits: magnetic resonance imaging (MRI), however, it only generates fluid velocity in Eulerian framework \citep{MRI}; similar limitation is seen in ultrasound imaging velocimetry (UIV) \citep{Gurung_ultra,Particle_Ultrasound_poelma}. Techniques such as positron emission particle tracking (PEPT) are capable of providing velocity information of radioactive tracers (single phase) in Eulerian and Lagrangian frames \citep{PEPT}, and have low spatial resolution compared to PIV systems. To the 
best of our knowledge, simultaneous solid-liquid velocity measurements for the transitional regime in particle laden flow are not 
available. Experiments of this nature with non-linear response to finite perturbations require a 
carefully controlled rig design with minimal experimental and ambient perturbation. Here we present 
a rig that has been specifically built to satisfy both sets of constraints. Whilst it elaborate on a 
number of techniques previously used in other rigs, the technical challenge resides in combining them
 into a single device.\\
In this context, we focus on Eulerian and Lagrangian velocity measurements conducted in a neutrally buoyant,
 particulate flow in a circular, horizontal pipe where both the experimental constraints imposed by 
transitional flows and those of two-phase velocity mapping are met.
The measurements are obtained by separating the dispersed and carrier 
phases using image brightness, and applying a combined 2D particle image/tracking velocimetry 
(PIV/PTV) to map the flow fields. The paper begins with a description of the two-phase facility, 
highlighting the design elements adopted from previous experiments for a precise control of experimental 
parameters (Section \ref{sec:1}); followed by presentation of the PIV/PTV measurement technique, 
including the steps involved in image processing and calibration to obtain the velocity fields 
(Section \ref{sec:3}).  Finally, velocity measurements for both the solid and fluid phases are presented in Sect. \ref{sec:7}.

\section{\label{sec:1}The two phase pipe facility}

\subsection{Description}
\label{sec:2}

\begin{figure*}
\centering
\includegraphics[width=\textwidth, keepaspectratio]{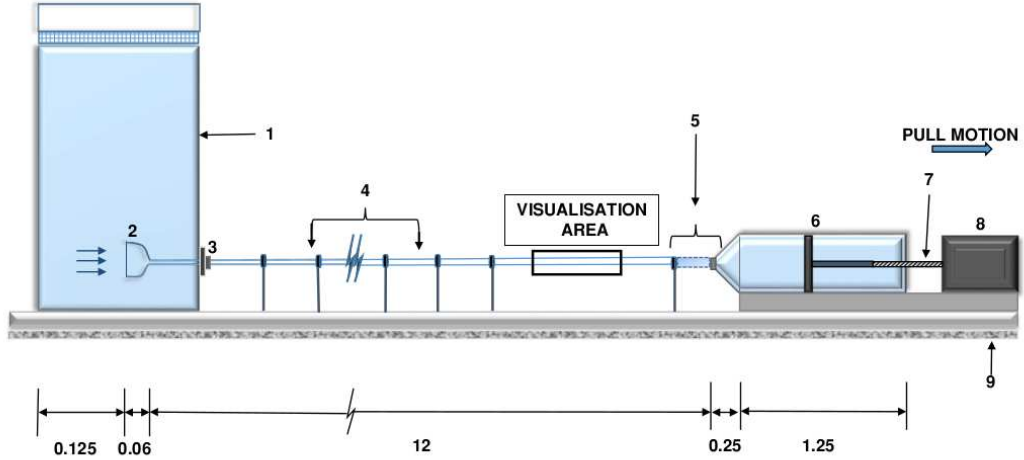}
\caption{A schematic diagram (front view) of the rig. (1) Reservoir, (2) Bell-mouth shaped inlet, (3) Flange, (4) Pipe sections held by unions on supports, (5) Flexible tube with a venturi, (6) Piston-Cylinder arrangement, (7) Lead screw, (8) Motor, (9) Adjustable Aluminum profile resting on concrete floor. All dimensions are in m. Drawing not to scale.}
\label{fig:1}
\end{figure*}

The rig consists of an assembly of glass pipes through which fluid and particles flow at a constant mass flux as shown in Fig.\ref{fig:1}. The reason for this choice is two-fold: First, this arrangement is capable of setting the flow velocity in a more controlled manner compared to pressure-driven systems \citep{darbyshire_mullin_1995}. Second, it avoids the sort of pressure fluctuations that prevents the accurate setting of flow rate \citep{Matas}. This point is crucial as we are interested in transitional flows, which are likely to be sensitive to finite amplitude perturbations. This constraint imposes a very controlled, and precise design such that the disturbances introduced from the experimental rig are kept to a minimum. For these reasons, the rig functions as an open loop: the pipe is fed by a header tank and both fluid and particles are pulled through it by a piston whose motion is controlled by a linear motor. We shall describe the details of its design, manufacture and performance. The key features of the facility are listed in Table \ref{tab:1}.
\paragraph{Pipe section and its assembly.}
The glass pipe has been assembled in 10 sections of cylindrical borosilicate glass tubes (GPE Scientific Ltd.). Each section is 1.2 m ($\pm$0.01 mm) long with an inner bore  diameter $D=20\pm0.01$ mm, and a wall thickness of 3.1 $\pm$0.03 mm. This configuration provides a ratio of the effective length of the pipe $L$ to diameter $D$, $L/D= 615$. Prior to their assembly, variation in concentricity of the pipe diameters were measured within a range of $\pm$10-30 $\mu$m by means of a 3-axis coordinate measuring machine, by mapping the maxima and minima of the inner and outer diameters of each pipe sections. This was taken as a tolerance limit for the manufacture of machined-to-fit Aluminum male/female flanges that served as unions between the pipe sections. The flanges were attached by pouring epoxy rubber to the pipe ends after carefully positioning them in a vertical jig on a rigid table. The jig consisted of a mandrel which ensured that both the pipe end, and the flanges were concentric, whereas the pourable rubber served to compensate for any expansion/contraction arising from temperature variations. The flange design and the assembly method were inspired from the Very Large Scale Pipe Flow (VLSPF) facility at the University of Liverpool \citep{DavidDenis}. 

Using selective assembly these pipe sections with unions attached were joined in a tight fit ranging from 10-30 $\mu$m. Male and female unions are separated by a rubber O-ring and PTFE tape to prevent any potential leak. To maintain the overall 
end-to-end straightness and horizontality of the pipe assembly, a target tool and a self-levelling LASER (Bosch) were used. The precision of the assembly is such that the LASER fired at one end of the long pipe assembly hits a target dot of 3 mm diameter within  $\pm$ 0.5 mm at the other end. The entire assembly rests on Aluminum profiles, with each union held by rubber reinforced clamps supported on adjustable threaded screws. This arrangement ensures that the overall horizontality of the entire rig is not affected by any slope in the floor, in addition to providing an isolation from vibrations emanating from the ground. Moreover, a gutter system is in place all along the rig to contain any potential spillage of the working fluid.
The pipe is fed by a rectangular header tank made of perspex with an inner volume of 62 l. A custom-made bell-mouth shaped inlet is attached to a shorter pipe section of 0.30 m in length. This inlet is placed at the center of the header tank (0.25 m$\times$0.25 m$\times$1 m) so as to minimise perturbations from the inlet walls, and joined to the rest of pipe sections through a flange on the tank wall. %
\paragraph{Working fluid and solid/liquid phases.} 
The liquid and solid phases are a critical element of the rig that must satisfy combined optical and mechanical constraints: First, the liquid must be transparent to allow for optical measurement 
techniques and the particles must be sufficiently reflective not only to be detectable, but also distinguishable from the tracers used to measure the fluid velocity measurements.\\
Second, fluid and particles must have matched density to cancel any buoyancy force. To satisfy both conditions, we chose an aqueous solution of Sodium Polytungstate (TC-Tungsten Compounds GmbH), obtained by diluting crystals in deionised water. 
The main advantage of this choice is that the fluid's density can be adjusted to match the density of the solid phase with an accuracy of 1 kg/m$^3$, simply by changing the dilution. 

For the purpose of our experiment, the solution has the following properties: density, $\rho_f$= 2500 Kg/m$^3$; dynamic viscosity, $\eta$ = 11-14 $\pm$0.001 mPa.s (at room temperature, T$_{r}$= 17$^\circ$-21$^\circ$C), and refractive index, \textit{n}= 1.52. These quantities are measured experimentally prior to each set of measurements. The density (Densitometer, Anton Parr) and the viscosity (Viscolite 700, Hydramotion) readings are temperature corrected measurements. The fluid is Newtonian with sufficient optical transparency for PIV techniques. 

However, the fluid undergoes a reduction reaction when in contact with any base metals (exception: Stainless steel, and some grade of Aluminum), resulting in discolouration which prevents optical measurements. To avoid this issue, extreme care has been taken in choosing the materials in such a way that the fluid only comes in direct contact with  glass, perspex, rubber material and anodised Aluminum parts. 

The density of the fluid has been adjusted to match that of standard spherical, monodisperse glass beads used for the solid phase (density $\rho_s$= 2500 Kg/m$^3$, \textit{n}= 1.50, Boud Minerals Ltd.), to achieve neutral buoyancy. 
The glass beads are graded using ISO standard sieves into required sizes (diameter, $d_p$ =150 $\mu$m $\pm$10 $\mu$m) and introduced in the fluid at a controlled volume fraction $C$ of $10^{-5}$. This simple methodology makes available to us monodisperse particles with a precisely controlled diameter in large enough quantities to fill the entire volume of the rig (62 l), at a reasonable cost. Furthermore, the close match between the refractive index of the fluid and glass beads could potentially allow measurements at higher volume fraction otherwise inaccessible to typical PIV measurements \citep{Wiederseiner2011}.
\begin{table}
\caption{Summary of the physical features of the rig.}
\label{tab:1}       
\begin{ruledtabular}
\begin{tabular}{lll}
Pipe & Diameter, $D$ & 20 mm \\
     & Length, $L$ & 12.3 m \\
     & Material& Borosilicate glass\\
     & Wall thickness & 3.1 mm\\
\hline\noalign{\smallskip}
Flow & Fluid phase & Sodium Polytungstate\\
     & Fluid density, $\rho_f$& 2500 Kg/m$^3$\\
     & Solid phase & Solid Glass beads\\
     & Glass bead density, $\rho_s$& 2500 Kg/m$^3$\\
\hline\noalign{\smallskip}
Seeding & Type & Silver-coated\\
        &       &hollow glass spheres\\
        & Density, $\rho_t$ &  2540 Kg/m$^3$\\
        & Diameter, $d_t$ & 32-38 $\mu$m\\
\end{tabular}
\end{ruledtabular}
\end{table}

Finally, for the purpose of optical measurements, the liquid is also seeded with smaller sized ($d_t$= 32-38 $\mu$m), silver-coated hollow glass microspheres (Cospheric) of nearly the same density, $\rho_t$= 2580 kg/m$^3$ as working fluid. These act as tracers passively following the fluid phase. The low settling velocity of the tracers ensures that they remain in suspension for hours.
\paragraph{Constant mass flux system.}
To ensure a constant mass flux, the fluid is drawn in by a system that consists of a modified pneumatic cylinder (Camozzi), 1 m in length and 0.20 m in inner diameter. The stainless steel piston of this cylinder is driven by an actuation system that is made of a leadscrew (IGUS, stroke length: 1.250 m, pitch: 5 mm), coupled to a DC motor (Oriental Motor, 200 W, Max. rated speed: 3000 RPM, Gear ratio: 20) with a dedicated speed controller (Oriental Motor, BLE2 series). The actuation system precisely sets the flowrate ($\pm$1 mm/s), which is also independently monitored by means of an Electromagnetic flowmeter (OMEGA, 5 to 250 l/min $\pm$ (1.5\% of reading + 0.3\% of range)). In addition, the rotation of the leadscrew is monitored using an encoder to ensure that the variations in linear velocity are well below 1\%.
\paragraph{Perturbation system.} 
To study the effects of finite amplitude perturbations on the stability of the flow, a perturbation 
generator has been devised that injects a small volume of fluid ($\sim$0.1\% of total mass flow) to the 
baseflow \citep{Peixinho,Hof_finite}. The system consists of an inlet (30 mm long, borosilicate glass inlet tube of inner bore of 2.2 mm) seamlessly attached on top mid-section of one of the pipe sections. The perturbation is a jet of fluid which is injected by a syringe coupled to a lead screw that is driven by a stepper motor. The fluid volume to be injected is extracted from the pipe during the syringe's retraction, and replenished in-between experimental runs. This is to ensure that the particle and tracer concentration remain constant. The jet is introduced at a distance of $225D$ downstream from the inlet. In all experiments, the flow is well established at that point.

\subsection{Control parameters}
The flow is determined by three non-dimensional control parameters: 
The Reynolds number $Re={UD}/{\nu}$, represents the ratio of inertia to viscous forces at the 
pipe scale in the fluid phase. Here, the average streamwise velocity of the fluid in the pipe is taken as $U$ where $U=u_x(r)$. The other two numbers characterize the particles and their interaction with the fluid phase: the Stokes 
number, $St={\tau_p}/{\tau_t}$, measures the ratio of 
the particle's relaxation time under the effect of drag, $\tau_p={\rho_p {d_p}^2}/{18\eta}$, and fluid's 
advection time $\tau_t={D}/{U}$. Lastly, the ratio of the total particle volume in the pipe to 
the working volume of the fluid yields the volume fraction of the particles $C$.
These parameters are independently controlled in the experiment, by controlling the total flowrate 
($Re$), particle diameter ($St$) and the amount of particles seeded in the flow ($C$). The range of 
accessible Reynolds numbers is determined by the minimum and maximum velocity of the piston and the 
maximum acquisition frequency of the optical system. In theory, the Stokes number can be adjusted to 
any value between micro-metric particles to spheres as large as the pipe allows. The concentration, on 
the other hand is limited by the reduction in optical visibility that occurs at high concentrations. This effect can be mitigated by the near-identical refractive indices of the fluid and the glass beads. However, 
our combined PIV-PTV system does not operate at concentrations greater than $0.1\%$. A summary of the 
range of physical and non-dimensional parameters is provided in Table \ref{tab:param}.
\begin{table}
\begin{ruledtabular}
\begin{tabular}{ll|rr}
$Re$ & 800-5060 & 0.21-1.33 & $U$ (m/s)\\
$St$ & 0.001-0.01 & 150-500  & $d_p$ ($\mu$m)\\
$C$ & $10^{-6}$-$10^{-3}$ \\
\end{tabular}
\caption{\label{tab:param} Range of Physical and non-dimensional control parameters currently accessible in the rig.}
\end{ruledtabular}
\end{table}
\section{Measurement system: PIV/PTV}
\label{sec:3} 
\subsection{Optical system}
\begin{figure}

\makebox[\textwidth][c]{\includegraphics*[trim= 310 280 200 60, clip, width=0.8\textwidth, keepaspectratio]{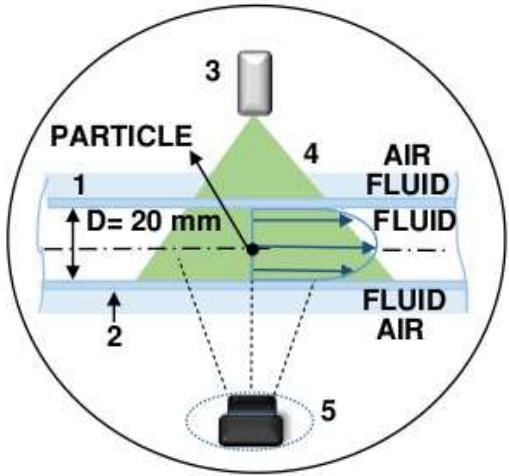}}

\caption{LASER measurement plane: longitudinal. (1) Visualization box, (2) Pipe, (3) LASER, (4) LASER sheet, (5) Camera.}
\label{fig:2}       
\end{figure}

As discussed in the introduction (see Sect.\ref{sec:intro}), simultaneous measurement of fluid/solid 
velocities in particle-laden flow is challenging due to presence of two sources of optical signals:
 tracers (used for the measurement of fluid velocity) and particles made of glass beads. Accurately 
separating the phases and measuring their respective velocities require a specific design of the 
optical measurement system. Consequently, a bespoke system 
has been devised to map the Eulerian and Lagrangian velocities of the carrier fluid and the solid 
particles. Eulerian velocities are obtained by PIV while Lagrangian velocities 
are obtained by PTV. For the measurement technique, a continuous wave (CW) 
LASER (OdicForce, 1 W) is used to generate a thin LASER light sheet (Thickness= 1 mm) with 
combination of a concave, a convex, and a cylindrical lens). Both phases are illuminated in a 
longitudinal plane, as shown in Fig.\ref{fig:2}. Successive images of the longitudinal plane 
are recorded by a camera (Photron Mini AX100) that can operate at 4000 Hz at a full resolution of $1,024\times1,024$ pixel, coupled with that of a macro 
lens (Tokina) with a shallow depth of field ($f-$number: 4). The measurement is conducted 180$D$ downstream from the point where perturbation is triggered. 
Some of the salient parameters of the PIV system are listed in Table \ref{tab:3}.\\ 
For an accurate measurement, the distortion of the images due to the curved walls of the pipe: hidden regions, multiple images and spurious displacements \citep{HofDoorne} have to be be avoided. To mitigate this, the measurement pipe section is immersed in a rectangular tank of 0.2 m length, 0.040 m height and 0.060 m depth filled with the working fluid. The density of the fluid in the visualization box is accurately matched to that inside the pipe. This minimizes refractive distortion.
\begin{table}
\begin{ruledtabular}

\begin{tabular}{lll}
Light sheet & LASER type & CW \\
     & Power & 1 W \\
     & Wavelength& 532 nm\\
     & Thickness & 1 mm\\
\hline\noalign{\smallskip}
Camera & Type & CMOS\\
     & Resolution & 1,024$\times$1,024\\
     & Sensor size & 20.48 $\times$ 20.48 mm\\
     & Pixel size &20 $\mu$m\\
     & Discretization& 8 bit\\
     & Frame rate & 2000-4000 Hz\\
     & Lens focal length & 100 mm\\
\hline\noalign{\smallskip}
Imaging & \textit{f}-number & 4\\
        & Viewing area & 0.027$\times$0.027 m$^2$\\
       & Primary Magnification \\
       &(PMAG) & 0.75X\\
\hline\noalign{\smallskip}
PIV&  &\\
analysis& Interogation area (IA) & 32$\times$32 px\\
    &Overlap IA& 75\% \\
     & Resolution (approx.) & 0.84$\times$0.84 mm$^2$ \\
     & Maximum tracer &12 pixel\\
     &displacement & \\
\hline\noalign{\smallskip}
PTV& &\\
& Average track length& 13 frame\\
& Bin size & 32$\times$32 px\\
%

\end{tabular}
\end{ruledtabular}
\caption{Some important parameters of the PIV system.}
\label{tab:3}       
\end{table}
\begin{figure}[ht!]
\centering
\includegraphics[scale=0.8, keepaspectratio]{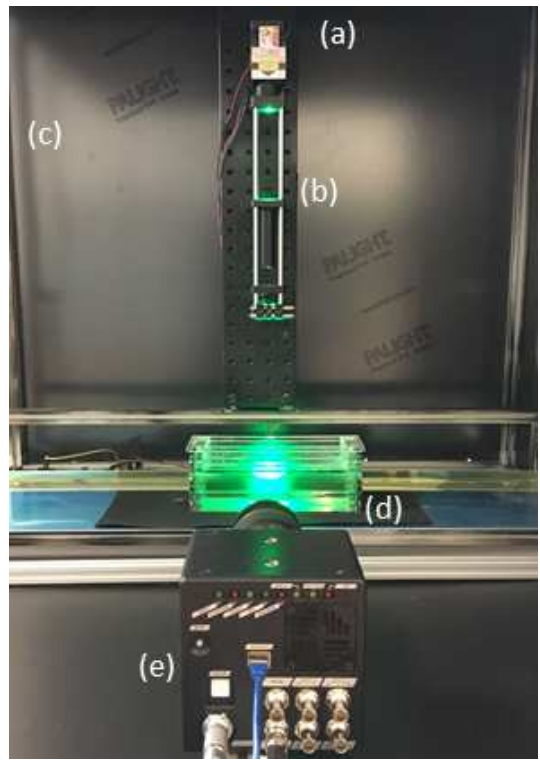}
\caption{Picture of the PIV setup. (a) LASER, (b) LASER sheet optics, (c) LASER box enclosure, (d) Visualization box, (e) Camera.}
\label{fig:3}       
\end{figure}
\subsection{Calibration.}
\label{sec:4}
Accurate calibration is required to translate distances measured in pixel in the image into actual 
physical distances. For this, a calibration tool of dog-bone type is devised. It consists of a rectangular section of width, 19.8 mm and length, 50.1 mm with a lattice of 49$\times$ 20 dots of 1 mm size spaced evenly by 1 mm. The rectangular section is attached to a circular section in the middle. This circular section is of the same width as the diameter of the pipe. The rectangular section is used for calibration in longitudinal direction. This tool is inserted from the inlet using flexible plastic rods, and accurately aligned with the LASER sheet in the 
measurement section. This ensures that the systematic error is minimal. Images of the calibration grid are taken which is then mapped onto a grid generated by \texttt{DaVis} software. A scale factor is then calculated which is later used to rescale the images such that the images are in world coordinates.
\subsection{Image Processing}
\label{sec:5}
We chose an example where the ratio of particle image diameter of tracer $d_\tau$ with that of solid particle, $d_s$ is $d_\tau$/${d_s}\approx$3.6. Due to the differences in sizes and optical properties between the particles and the tracers 
\citep{kiger_pan_2000}, a phase discrimination method is first applied to detect, and separate the two 
phases using a tracking algorithm adapted from the work by \cite{Kelley}. 
Following the separation, the PIV \& PTV techniques can be applied. The image processing involves the 
following steps as shown in the flowchart (Fig.\ref{fig:12}):
\begin{figure*}
\makebox[\textwidth][c]{\includegraphics*[trim= 0 0 0 0, clip, width=\textwidth, keepaspectratio]{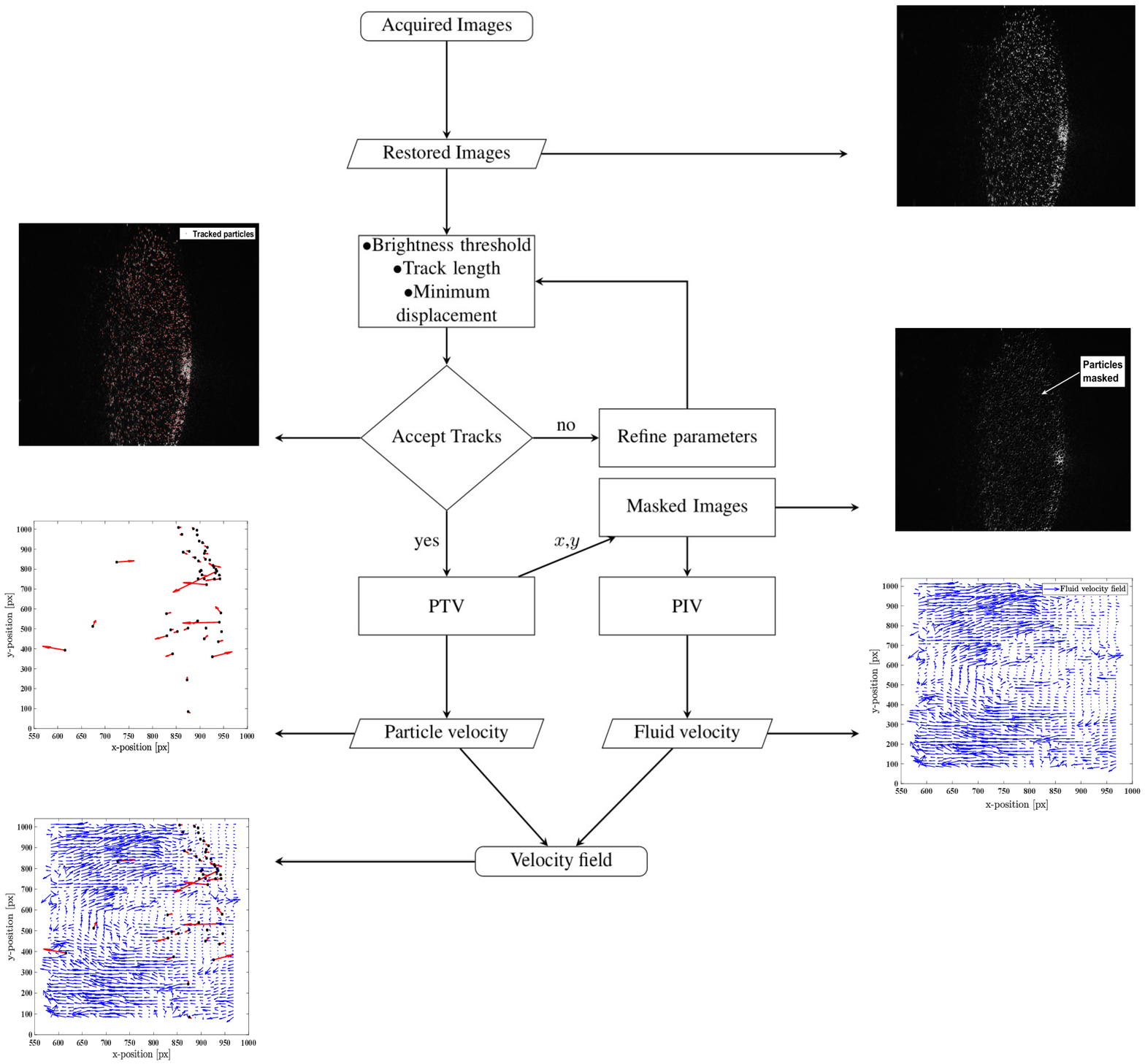}}
\caption{Schematic representation of the procedure to measure both the Lagrangian velocity of particles and the Eulerian fluid velocities from the same set of images. The axes of the figures are in pixel [px].}
\label{fig:12}
\end{figure*}
\begin{enumerate}
\item Removal of image distortions arising from background noise, uneven illumination and 
specular reflections; the steady background images are removed by subtracting the global pixel mean of all images; 

\item Detection of bright particles (solid \& tracers) by setting a brightness threshold \emph{i.e.}, a 
local high intensity maxima in the restored images, only the brightest ``blobs'' that differ from the 
background contrast by a prescribed brightness threshold are retained. This optimal threshold is chosen by plotting a gray-level histogram of the 21K images with the object average intensity $I$ defined as brightness in $x$-axis, and number of pixels $Npx$ in the $y$-axis. To illustrate the manner in which the parameter used to segregate the background from objects are chosen, a gray-level histogram from a two-phase experiment is shown in Fig. \ref{fig:PTV_threshold}. A \emph{bimodal} histogram is evident from the two distinct histograms of objects (tracers and solid particles) and background. The optimum threshold parameter is the point where the curves from the histogram fit intersect at a gray-level of 157 in the $x$-axis. Since the particle concentration is low, the particle distribution peak is not distinctly visible.
\begin{figure}
\centering
\includegraphics[width=0.5\textwidth,keepaspectratio]{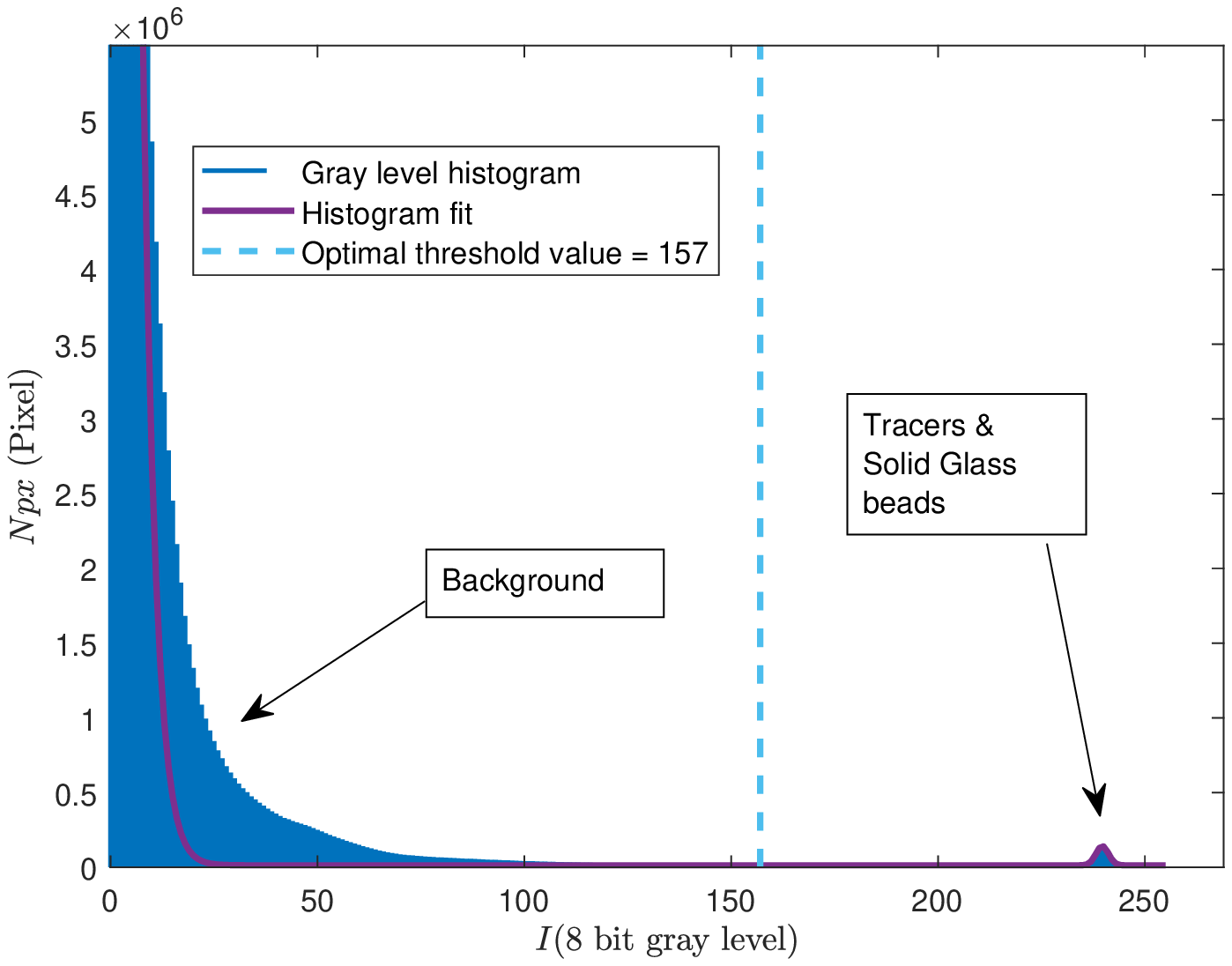}
\caption{Threshold for background separation. Gray-level histogram of the images with the object average intensity $I$ defined as brightness, and $Npx$ as the number of pixels. Two distinct histograms are seen.}
\label{fig:PTV_threshold}
\end{figure}
Once the background is cut-off, the tracking algorithm is set to search for gray pixel areas greater than 1 pixel. On plotting the size-brightness map of detected particle from the search, separation limits can be distinctly seen as shown in Fig.\ref{Fig:PTV_parameters} where regions containing the tracer and the particles are visible. The size-brightness map represents the total amount of signal of the particles assuming it to be the product of particle size $S$, its brightness as maximum intensity $I_{max}$ and number of particles detected $N$ with the log of this product represented by the color scale is in \cite{Khalitov2002}. The images were not pre-treated using any filter and so appear to be slightly noisy.

\begin{figure}
\centering  
\includegraphics[width=0.5\textwidth,keepaspectratio]{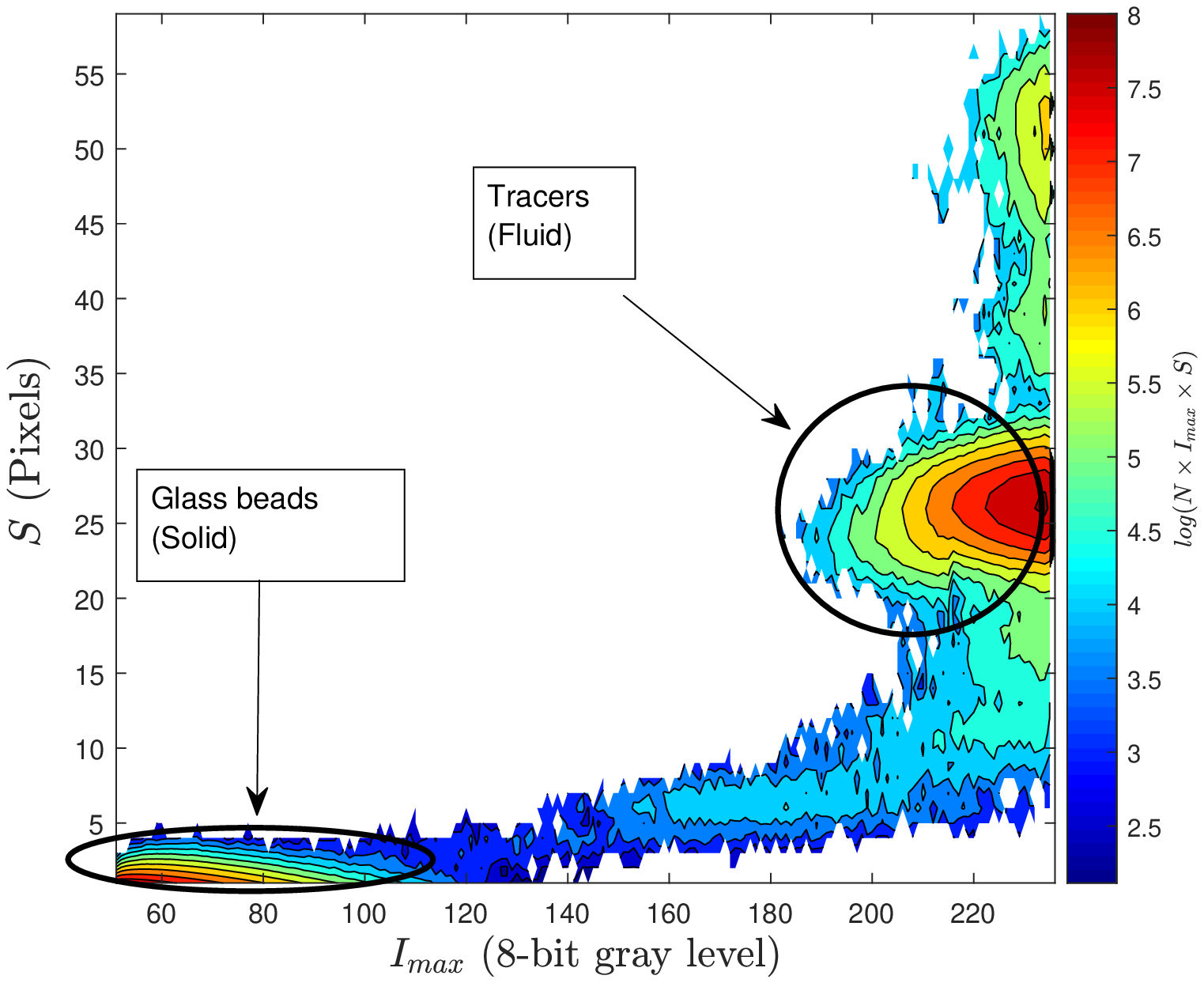}
\caption{Phase separation limit of regions of tracers and glass beads as illustrated by a size-brightness map consisting of the product of particle size $S$, its brightness as maximum intensity $I_{max}$ and number of particles detected $N$. The logarithm of this product represents the color scale.}
\label{Fig:PTV_parameters}
\end{figure}
A clear separation limits of the tracer and particle intensities is seen, illustrating the capacity of the tracking algorithm to successfully separate phases by choosing particles set by the parameters. The location of the centers of solid particles is determined by fitting two 1D Gaussian Estimator to the adjacent pixels, one for the horizontal position, the other for the vertical position. Since the intensity profile of particle image is not known, Gaussian profile is commonly assumed. This results in center estimation with a sub-pixel accuracy \citep{Ouellette2006};
\item The solid particles detected in step (2) are accepted for tracking only if they are within the maximum displacement threshold \emph{i.e.}, a search window within which a detected particle is allowed to move in the subsequent image. This generates an estimate for the measured particle position in $x$ and $y$ in each image. The tracks of detected particles are matched using a predictive three-frame best-estimate algorithm. This algorithm initially takes two images and uses nearest neighbor search to initiate linking of tracks. The tracks in subsequent frames are approximated using three image frames. Assuming least change in velocity in subsequent frames as the optimal solution, particles that remain within the specified search window are accepted as valid Lagrangian data. The Lagrangian data is then differentiated in time to generate time series of positions and velocities of tracked solid particles. The algorithm computes the time derivatives by convolution with a Gaussian smoothing and differentiating kernel using the method proposed by \cite{MORDANT2004245}. An optimum choice of the filter width, $w$ and the fit width, $T$ was made by varying the two parameters and observing the length of tracks, the number of particle detected, and the RMS error. $w$= 1 and $T$= 2.5$w$ was chosen after the parametric tests.

This algorithm handles overlapping particles by ending tracks at point of overlap, and starting new tracks. This increases the data yield compared to an algorithm which cancels conflicting tracks. More algorithmic details of the tracking code can be found here \citep{Ouellette2006}.

\item Finally, the liquid phase is separated from the solid phase by dynamic masking. Logical masks using the $x$ and $y$ co-ordinates of the particles detected in step (3) in each frame are set to the background image pixel, hence keeping the images with tracers only. The particles are masked by circles of the same size as the particles. These circles are generated using a MATLAB function, and filled with pixels of the same grey level as the background. In addition, the masked images are pretreated by a 3$\times$3 median filter to remove any leftover traces of ``coronas'' of partially masked during pre-processing \citep{brucker2000piv,minier2016particles}. Finally, the images are processed using standard PIV data analysis software (\texttt{DaVis} 10) to obtain the fluid velocity field. During the \texttt{DaVis} processing, the images are cleared of any specular reflections by background subtraction. This step is followed by PIV processing wherein the images are cross-correlated in 4 passes starting with 128$\times$128 px interrogation window with 50\% overlap, and ending in a final pass of a 32$\times$32 px window with an overlap of 75\%. The vectors with a dispersion of more than 2.5 standard deviation are iteratively removed and replaced. Using the methodology proposed by \cite{Sciacchitano_2016}, \texttt{DaVis} computes the PIV uncertainty quantification of statistical quantities. To perform advanced statistics, the PIVMat Toolbox is used. 
\end{enumerate}
\subsection{Experimental procedure}
First, the fluid is mixed with tracers and glass beads, and stirred within the header tank. The fluid properties are then measured and recorded. The piston is first activated in alternative directions few times so as to homogeneously mix the suspension. Both particles and tracers 
remain in suspension for hours due to their practically zero settling velocity. There is an idling time 
of about 20 min before every experimental run to allow the fluid to become quiescent. This process is 
determined empirically, the idling time is longer after runs at higher $Re$.\\
For our measurements, the Reynolds number was in the range $Re$= 800-5060 and the solid 
particle concentration was set to $C$=$10^{-5}$. The LASER sheet was oriented in the longitudinal plane and 
the viewing area was approximately $0.027\times0.027$ m$^2$. The PIV images were taken during the pull motion of the piston. The image acquisition was triggered typically after 20 s of experimental run, and lasted for 5-10 s. The images are saved in a hard disk drive (HDD) in full depth, Raw image format and later exported as 8-bit TIFF files for image processing. A typical acquisition generates a dataset of 39 GB, consisting of 21,841 images.
\subsection{Validation of the processing technique on synthetic images}
\begin{figure}
\includegraphics[width=0.3\textwidth, keepaspectratio]{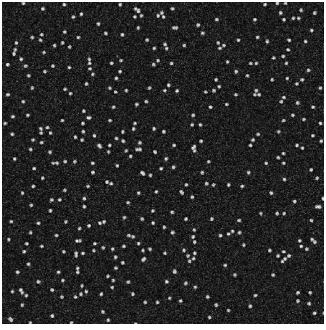}
\caption{A sample of cropped synthetic image (450$\times$450 px) generated by \texttt{DaVis} is shown.}
\label{fig:13}      
\end{figure}

\begin{figure}[h!]
\includegraphics[width=0.45\textwidth, keepaspectratio]{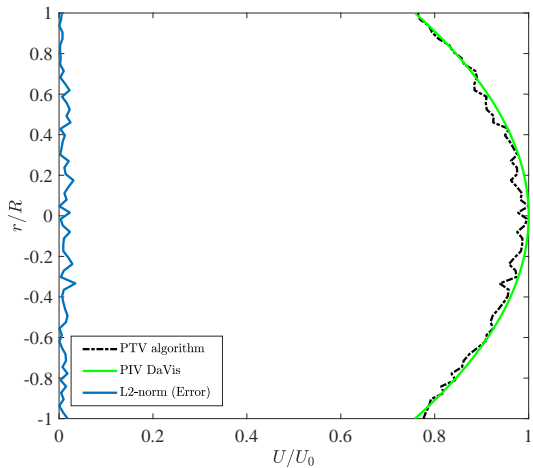}
\caption{Profiles of streamwise velocity, $U$ scaled by $U_0$ against radial distance $r$ scaled by pipe radius $R$, generated using the reference velocity field from \texttt{DaVis}, and by the PTV algorithm respectively. The RMS of the error is shown as well.}
\label{fig:14}      
\end{figure}

To estimate the PTV tracking error, synthetic images were used. \texttt{DaVis} was used to generate 4096 synthetic images of 1024 $\times$1042 px with randomly distributed particles of 2 px in sizes. The particles had a Gaussian intensity profile, and the seeding density was 0.001 particles per pixel (ppp). In addition, an additive Gaussian noise (25\%) was added to the images, and the particles were given a background laminar flow (shift of 5-10 px) for which a reference velocity field was generated. A sample synthetic image is shown in Fig. \ref{fig:13}.
The PTV tracking algorithm was run on these synthetic images. Taking the reference velocity field as the ideal case, the discrepancies in velocity field generated by the PTV was computed. The RMS of the error is plotted in Fig. \ref{fig:14}, along with the reference streamwise velocity profile, and the streamwise velocity profile using PTV algorithm respectively. The velocities are normalized by the respective centreline velocity. 
The tracking algorithm follows the reference velocity profile with a maximum error of 0.34 pixel. The slight non-smoothness of the PTV profile is characteristic of Lagrangian data when they are averaged over smaller bins. This is typically improved by averaging over wider bins, or by gathering more data points. In real experiments, the PTV algorithm is run on 21k images; suggesting that the error is further minimized. 
\section{Profiles of fluid and particles velocities}
\label{sec:7}
\subsection{Single phase flow}
\label{sec:8}
\begin{figure}
\includegraphics[width=0.5\textwidth, keepaspectratio]{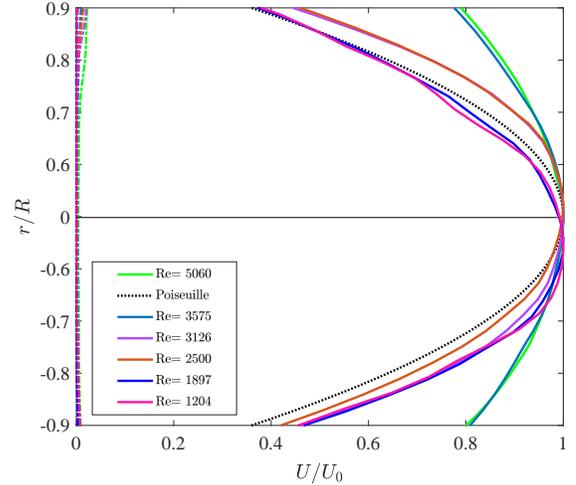}
\caption{The time-averaged streamwise velocity $U$ profile (scaled by centreline velocity $U_0$) against radial distance $r$ (scaled by pipe radius $R$) is shown for $Re$= 1200-5060 with their corresponding RMS fluctuations. Theoretical Poiseuille flow is plotted for reference.}
\label{fig:6}       
\end{figure}

\begin{figure}

\includegraphics[width=0.6\textwidth, keepaspectratio]{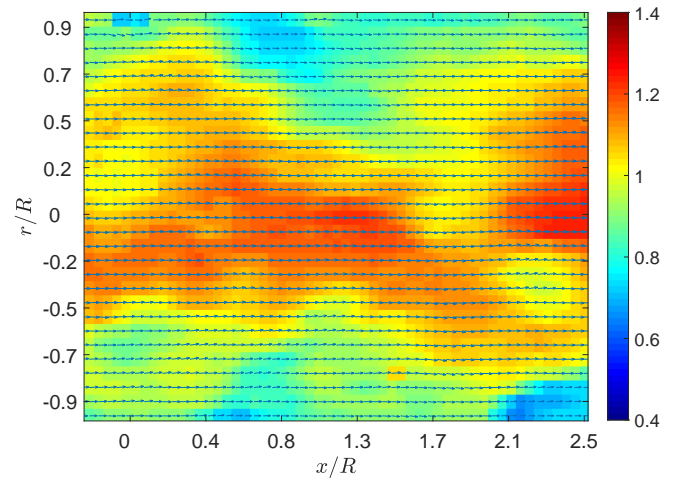}

\caption{A snapshot of single phase flow field with its velocity norm scaled by $U$ at $Re$= 3575 shows regions of turbulence (Multimedia view 1). Instantaneous velocities are shown in arrows.}
\label{fig:15}       
\end{figure}

\begin{figure}
\includegraphics[width=0.45\textwidth, keepaspectratio]{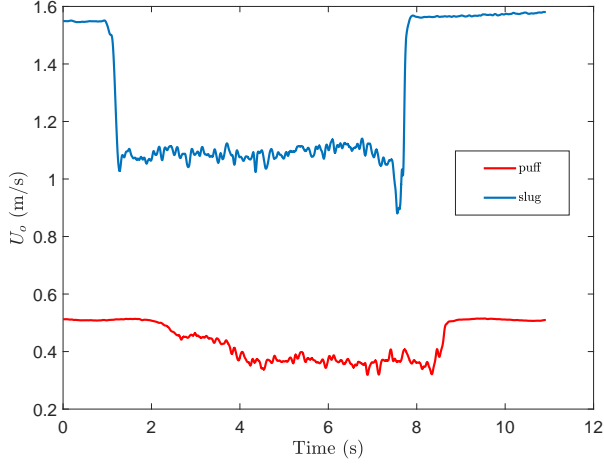}
\caption{Streamwise velocity time trace for single phase flow probed at the centreline $U_o$ for $Re = 800$ (red curve) showing a puff, and $Re = 3400$ (blue curve) showing a slug.}
\label{fig:18}  
\end{figure}

The pipe flow was first characterized by a single phase experiment with the heavy fluid seeded with 
tracers only. PIV measurements were performed in the longitudinal plane, for values of Reynolds spanning 
$Re$= 1204-5060, so as to cover both laminar and turbulent regimes. No perturbation was applied.
The mean streamwise velocity measurement obtained is shown in Fig.\ref{fig:6}. For $Re\leq2800$, the 
time-averaged streamwise component of the fluid velocity, $U$, shows 
a typical Hagen-Poiseuille parabolic flow profile, in good agreement with the expected laminar 
theoretical profile. The profiles with $Re$= 3575 and $Re$= 5060 show a flatter profile, closer to what 
is expected from a turbulent flow. 
Inspection of the time-dependent flow-field reveals that intermittent bursts of turbulence are present 
in this regime, as expected in the transitional regimes. A snapshot of velocity norm scaled by $U$ is shown in Fig. \ref{fig:15} (Multimedia view 1) for $Re$= 3575.\\
Using the perturbation system, signatures of puff and slug can be seen in the velocity-time trace as shown in the Fig.\ref{fig:18}. For an impulsive jet of 0.75 ml with an injection time duration of 43 ms, a slug (blue) was seen at $Re = 3400$, and a puff (red) was observed at $Re=800$. To obtain this, raw velocity data was smoothed by a Gaussian window (window length= $0.1\%$ of the total frame number) chosen empirically. Following that, streamwise velocity at centreline of the pipe is extracted. Since the measurements are done at one location, it is hard to conclusively categorize the flow structures, and infer if the appearance of puff and slug would lead to sustained transition eventually. However, their presence can be directly measured.

\subsection{Two-phase flow}
\label{sec:9}
\begin{figure}
\includegraphics[width=0.5\textwidth, keepaspectratio]{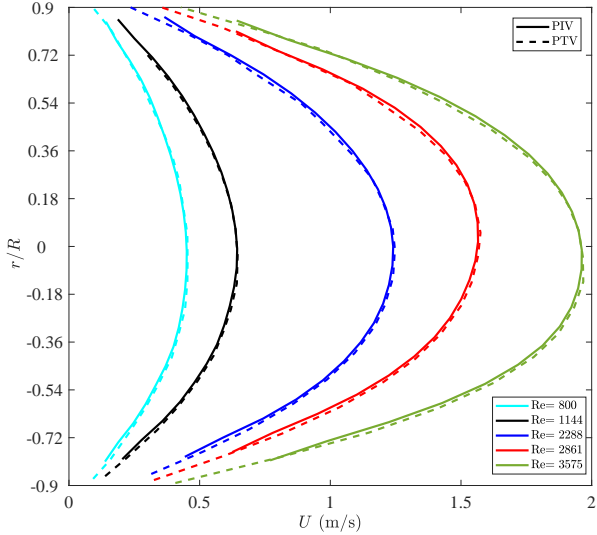}
\caption{Time-averaged streamwise velocity $U$ profiles against radial distance $r$ (scaled by pipe radius $R$) for solid (dashed line) and fluid phases (solid line).}
\label{fig:11}      
\end{figure}
\begin{figure}
\includegraphics[width=0.55\textwidth, keepaspectratio]{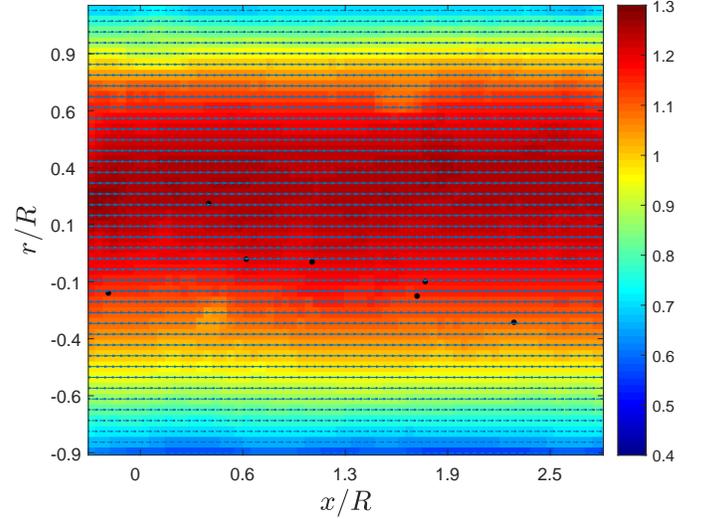}
\caption{A snapshot of the two-phase flow field with its velocity norm scaled by $U$ at $Re$= 3575 shows high centreline velocity typical of laminar flows. Instantaneous velocities are shown in arrows while the particles are shown as black dots (Multimedia view 2).}
\label{fig:16}       
\end{figure}
\begin{figure}
\includegraphics[width=0.45\textwidth, keepaspectratio]{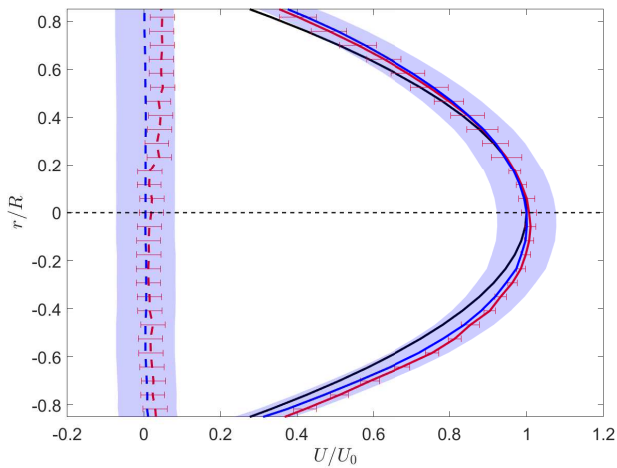}
\caption{Measurement uncertainties for PIV and PTV measurements at $Re$= 2000. The time-averaged streamwise velocity $U$ profile (scaled by centerline velocity $U_o$) against radial distance $r$ (scaled by pipe radius $R$). Shaded area shows uncertainty for PIV, whereas the error bars are for PTV. Theoretical Poiseuille flow (black curve) and the pipe centerline (dotted black line) are shown as well.}
\label{fig:17}       
\end{figure}
\begin{figure}
\centering
\includegraphics[width=0.45\textwidth]{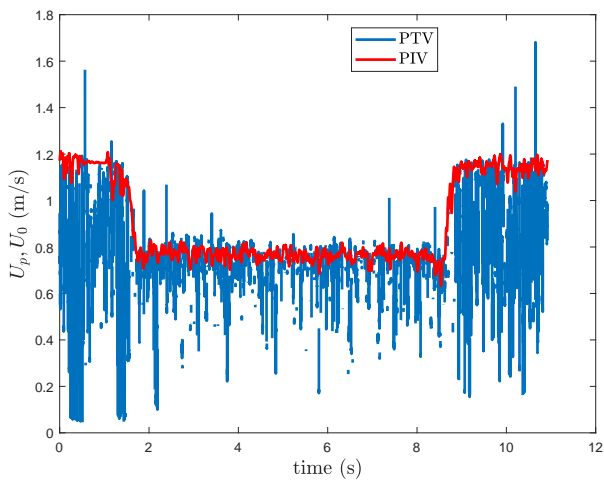}
\caption{Streamwise velocity time trace for a two phase flow at $Re = 2500$. Eulerian fluid velocity is shown in red (probed at centreline $U_0$), whereas  the Lagrangian particle velocity $U_p$ averaged over specified window is shown in blue.}
\label{fig:19}
\end{figure}

Next, we investigated a two-phase flow at low concentration $C$= $10^{-5}$, $St$= 0.01-0.001 for a 
range of Reynolds numbers spanning $Re$= 800-3575. The average velocity profiles of both the fluid, and the solid phases are shown in Fig.\ref{fig:11}. Both are very close to the laminar Poiseuille profile. This result is expected since at the current low particle load factor, the influence of the solid particles on the flow is negligible. Hence, the fluid and the particles are only coupled one-way, with the solid particles acting as tracers \citep{Matas,anthony}. A snapshot of the velocity norm scaled by $U$ of the particulate flow at $Re$= 3575 is shown in Fig. \ref{fig:16} (Multimedia view 2). Both the fluid and the particles follow the same laminar streamlines suggesting that the flow is laminar at $Re$ = 3575, unlike the single phase at the same Reynolds number. 
 The uncertainty in the time-averaged fluid velocity measurements, based on the discrepancy from a perfect cross-correlation (obtained from \texttt{DaVis}) was found to be within a range of 5-8\% (mean velocities). The uncertainty in PTV measurement is computed using the RMS. Fig.\ref{fig:17} shows the typical uncertainties in time-averaged streamwise velocity with their corresponding RMS fluctuations for PIV and PTV measurements. For reference, these are only shown for measurement at $Re$=2000 with the order of uncertainties remaining the same for measurements at other Reynolds numbers. 
As discussed in \ref{sec:8}, puffs and slugs can be triggered by injecting impulsive jets. However, tracing them using Lagrangian velocity field is more complex than its single phase counterpart. While centerline velocity in Eulerian velocity field is extracted by simply positioning a ``probe", this is not possible in Lagrangian data due to sparseness of such data. Instead, the discrete velocity data are averaged over a window of a specified size positioned at the centreline of the pipe. All particles detected in this window account for the average streamwise velocity of the particles $U_p$. Figure \ref{fig:19} shows velocity time trace of a slug at $Re$= 2500 using Eulerian (red) and Lagrangian (blue) velocities. The time evolution of particle velocities is noisy because data are collected from a large averaging window spanning areas of low velocity near the wall and of high velocity near the centreline.
However, when overlapped with the time evolution of the fluid velocity (probed at the centreline $U_0$ such that $U_0=u_x(r=0,t)$), the fronts in the particle velocity trace can be clearly observed.

\section{Summary and conclusions}
\label{sec:10}
We have presented a unique pipe flow facility with a novel optical measurement system, built in the express purpose of studying the transition to turbulence in particulate pipe flows. To this end, we have presented the first-ever simultaneous Eulerian-Lagrangian fluid-particle velocity measurements for very dilute ($C$= $10^{-5}$), particulate pipe flow undergoing 
transition to turbulence within a range of $Re\in[800-3575]$. In addition, we have also demonstrated how image separation techniques, and an existing PTV tracking algorithm can be adapted to form a novel combined 2D PIV/ PTV 
technique. The masking and phase discrimination method we implemented is capable of successfully separating the phases. Moreover, the PTV algorithm 
performs well with negligible tracking error. The experimental facility was shown to 
retain an undisturbed Poiseuille flow for $Re$ up to 3000. On the other hand, the 
PIV/ PTV measurement system effectively captures known particle dynamics at low $C$, and is adept at characterizing transition by successfully tracing transitional flow patterns in both solid and liquid phases. These observations 
provide a validation of the experimental facility and the measurement system respectively for the 
future characterization of transitional particulate pipe flows.\\
The present technique nevertheless has certain limitations: it does not work with high particle 
concentrations when particles obscures each other such that the linking of PTV tracks become erroneous 
(particle concentration of the order of $10^{-2}$). Additionally, trajectories linking is possible only 
when single time step displacements of particles are reasonably small. This can be well mitigated by
means of a high-speed camera. Similarly, for the PIV, images with more masked regions will render false 
velocity correlations. In principle, in planar PIV measurements these challenges can only be addressed 
to a certain extent by careful refractive index matching, or by adopting combinations of other existing 
algorithms 
\citep{Brevis2011,Dalziel1992,SBALZARINI}. For higher concentration 
measurements, 3D PIV/PTV or advanced techniques such as positron emission particle tracking have to be 
employed. However, our research does not aim to look at the effects of higher particle 
concentration. Given the various design and measurement challenges, this is the only measurement system that can successfully reconcile all of it. To conclude, the present experimental facility with its unique measurement system is well suited to address 
the question of how a dilute-to-moderate concentration of particles affects the transition in pipe flows.   
\begin{acknowledgements}
AP is supported by a Wolfson Research Merit Award from the Royal Society (grant WM140032)). CCTP is partly supported by EPSRC grant No. EP/P021352/1. We thank the Environmental Complexity Lab, Stanford University for making the Particle Tracking code freely available online.
\end{acknowledgements}

\bibliography{aipsamp}

\end{document}